\documentclass[10pt]{article}

\advance \textheight by \topskip
\usepackage{caption}
\usepackage{amssymb,dsfont,amsmath,amsfonts,bm,mathabx}
\usepackage{slashed}
\usepackage{color}
\usepackage{graphicx}
\usepackage{cite}
\usepackage{tikz}

 \def\*{{\varstar}}

\def\be{\begin{equation}}
\def\ee{\end{equation}}
\def\bea{\begin{eqnarray}}
\def\eea{\end{eqnarray}}

\def\a{\alpha}
\def\b{\beta}

\def\bfa{{\boldsymbol{\alpha}}}
\def\bfb{{\boldsymbol{\beta}}}

\def\bfxi{{\boldsymbol{\xi}}}
\def\bfz{{\boldsymbol{\zeta}}}

\def\bmu{{\boldsymbol{\mu}}}
\def\bnu{{\boldsymbol{\nu}}}
\def\blambda{{\boldsymbol{\lambda}}}
\def\brho{{\boldsymbol{\rho}}}

\def\bPhi{{\boldsymbol{\Phi}}}
\def\bPsi{{\boldsymbol{\Psi}}}

\def \bY{{\bf Y}}
\def\bW{{\bf W}}

\begin{document}

\title{Higher spin matrix models}

\author{\sf  Mauricio Valenzuela\footnote{\texttt{valenzuela.u} \textit{at} \texttt{gmail.com}}
\\[5mm]
{\textit{Facultad de Ingenier\'ia y Tecnolog\'ia}}\\
{\textit{Universidad San Sebasti\'an, General Lagos 1163, Valdivia 5110693, Chile}}
}
\date{}

\maketitle

\begin{abstract}
 
We propose a hybrid class of theories for higher spin gravity and matrix models, i.e. which handle simultaneously higher spin gravity fields and matrix models. The construction is similar to Vasiliev's higher spin gravity but part of the equations of motion are provided by the action principle of a matrix model. In particular we construct a higher spin (gravity) matrix model related to type IIB matrix models/string theory which have a well defined classical limit, and which is compatible with higher spin gravity in $AdS$ space. As it has been suggested that  higher spin gravity should be related to string theory in a high energy (tensionless) regime,  and therefore to  M-Theory, we expect that our construction will be useful to explore concrete connections.

\end{abstract}

\section{Introduction}

 Vasiliev's higher spin gravity ({\bf HSGR}) is a gauge theory whose spectrum contains an infinite tower of massless higher spin and matter fields which enjoy infinite dimensional higher spin symmetries. In the tensionless (high-energy) limit  of string theory its spectrum is spanned by massless modes which exhibits higher spin symmetries { \cite{Gross:1988ue}  (also in the first quantized level \cite{Engquist:2007pr}), expectedly equivalent to those of HSGR \cite{Vasiliev:2001ur}}. The latter arguments have led to the idea that higher spin gravity might describe a high-energy/tensionless-limit of string theory (see e.g. { \cite{Sundborg:2000wp,Sezgin:2002rt,Bonelli:2003kh,Engquist:2005yt,Fotopoulos:2010ay,Sagnotti:2013bha}}), or reciprocally that string theory might appear as a broken symmetry phase of some version of  higher spin gravity. This also suggests that HSGR is a manifestation of M-theory.

As for matrix models ({\bf MM}s),  it has been argued that M-theory should be described by a matrix model  \cite{Banks:1996vh,Susskind:1997cw} in $0+1$ dimensions. It has been also conjectured in \cite{Ishibashi:1996xs} that certain matrix model in zero dimension should describe an effective theory of type IIB strings. Both statements suggest that matrix models might be useful for the description of non-perturbative aspects of strings and M-theory. 

In this paper we show that Vasiliev's HSGR in $D$ spacetime dimensions can be regarded as a type of relativistic MM in $D$ dimensions and we shall argue that certain modification of Vasiliev's equations which incorporate the action principle of matrix models will permit the construction of a new type of higher spin gravities.  The latter, dubbed here {\it higher spin matrix models}, will be straightforwardly related to matrix models. In particular we shall construct the type IIB higher spin matrix models which combines  vanishing (higher spin gravity) curvature and constant (matter fields) covariant derivatives conditions with type IIB matrix model equations of motion \cite{Fayyazuddin:1997yf}.

\section{Matrix models and higher spin gravity} 

There is an interesting connection between super-Yang-Mills and string theories in ten dimensions proposed in \cite{Ishibashi:1996xs}. The pure interaction part of super-Yang-Mills, extended with a chemical potential term, obtained after compactification of all directions to size zero is given by the MM action,
\be \label{MM1}
S_{\tt MM1}=\left\langle \alpha \left( -\frac{1}{4} [A^I, A^J] [A^I, A_J] - \frac{1}{2} \bar{\psi}\gamma^I[A_I, \psi]\right) +\beta \mathds{1} \right\rangle, \quad I,J = 0,1,...,D-1.  
\ee
Here $\langle \cdot \rangle$ is the (super)trace of the theory, which respects the correspondent cyclic properties for boson ($A$) and fermion ($\psi$) fields and $\a$ and $\b$ are some constants.  When the size of the matrices $A$ and $\psi$ goes to infinity they may be regarded also as Hilbert space operators and therefore they must possess a classical limit, where they become functions on a classical phase-space and their commutator product will become a Poisson bracket. The classical limit of the action \eqref{MM1} can be identified with a string action treating the worldsheet as a phase-space and the fields $A$ and $\psi$   as functions on this phase-space.

\begin{table}[h]
\begin{center}
\begin{tabular}{|c|}\hline
\hbox{IKKT matrix model}\\ \hline
gauge field: $A^I$  \\[4pt]
commutator: $[\cdot, \cdot] $ \\[4pt] 
(super)trace: $\langle \cdot \rangle$ \\[4pt] 
$S_{\tt MM1}$\\
\hline
\end{tabular}
\quad 
$-$ Classical limit $\rightarrow$
\quad 
\begin{tabular}{|c|}\hline
\hbox{String theory} \\ \hline
target space coordinate: $X^I$  \\[4pt] 
Worldsheet Poisson bracket:  $[\cdot, \cdot]_P $ \\[4pt] 
Integral operator: $\int d^2\sigma \, \sqrt{-g(\sigma)}$ \\[4pt] 
$S_{\tt Sch}[Y,\psi , g ]$\\
\hline
\end{tabular}
\end{center}
\caption*{From IKKT model to type IIB string action.}
\end{table}

The conjecture of the authors \cite{Ishibashi:1996xs} says that their (IKKT) matrix model is related to the type IIB Green-Schwarz action in the Schild gauge \cite{Schild:1976vq,Ishibashi:1996xs},
\be\label{Sch}
S_{\tt Sch}[Y,\psi , g ] = \int  d^2\sigma  \,\left(\alpha \left( \frac{1}{4\sqrt{-g(\sigma)}} \, [Y^I,Y^J]_P [Y_I, Y_J ]_P -\frac{i}{2} \bar{\psi}\gamma^I [Y_J\,, \psi]_P\right) + \beta \sqrt{-g(\sigma)} \right) , 
\ee
where { $Y^I $} are the spacetime coordinate components of the string and $\psi$ is an (anti-commuting) Majorana spinor. Here the Poisson bracket is defined as,
\be \nonumber
[X,Y]_P=\frac{\partial X}{\partial\sigma^1} \frac{\partial Y}{\partial\sigma^2}- \frac{\partial X}{\partial\sigma^2} \frac{\partial Y}{\partial\sigma^1},
\ee
and $\sqrt{-g(\sigma)}$ is the determinant of the induced metric on the worldsheet, which can be regarded also as an independent scalar field. 

The equations of motion obtained from \eqref{Sch} are
\be \nonumber
[Y^J, [Y_I, Y_J]_{P}]_P=0\,, \qquad \gamma_I [Y^I\,, \psi]_P=0,
\ee
and from the the variation of $\sqrt{-g(\sigma)}$,
\be \label{eomclass2}
-\frac{\alpha}{4\sqrt{-g(\sigma)}^2}[Y^I,Y^J]_P [Y_I, Y_J]_P+\beta=0.
\ee
Note that the current $\bar{\psi}	\gamma_I \psi$ vanishes owing to anti-symmetry of the fermion product $\psi_\a\psi_\b$ and the symmetry of the matrices $(C\gamma^I)^{\a \b}=(C\gamma^I)^{\b \a}$, where  $C$ is the conjugation matrix, and $\gamma_I$ are in the Majorana representation.
From \eqref{eomclass2} we can solve for $\sqrt{-g(\sigma)}$, and replacing back its value in the action \eqref{Sch} it returns the Nambu-Goto form (see e.g. \cite{Makeenko:1997zb}),
\be\nonumber
S_{\tt NG}[Y,\psi] = {\tt sign}(\beta) \sqrt{2 |\a | |\b | } \int  d^2\sigma  \,\left(\sqrt{\frac{{\tt sign}(\a\beta)}{2} [Y^I,Y^J]_P [Y_I, Y_J]_P} - i {\tt sign}(\a\beta)  \frac{1}{2}\sqrt{\frac{|\a |}{2| \b |}}  \bar{\psi}\gamma^I [Y_I\,, \psi]_P\right), 
\ee
which is equivalent to
\be\label{SNG}
S_{\tt NG}[Y,\psi] = -T  \int  d^2\sigma  \,\left(\sqrt{- \frac{1}{2} [Y^I,Y^J]_P [Y_I, Y_J]_P} +2i \bar{\psi}\gamma^I [Y_I\,, \psi]_P\right)  , 
\ee
for
\be \nonumber
 {\tt sign}(\beta)=-1, \quad T=\sqrt{2 |\a | |\b |  },\quad {\tt sign}(\a\b)=-1,\quad \sqrt{\frac{|\a |}{2| \b |}}=4.
\ee
This means, 
\be \label{alphabeta}
\a ={4T},\quad \b =-\frac{T}{8}.
\ee
Note  that in a flat background the Poisson bracket is equivalent to the determinant of the induced metric on the workdsheet,
\be \nonumber
{\tt det}(g_{IJ} \partial_\alpha Y^I \partial_\beta Y^J)=\frac{1}{2} [Y^I,Y^J]_P [Y^I,Y^J]_P,\quad \a,\b=1,2,
\ee
so that \eqref{SNG} can be turned int the standard form found in many textbooks.

A variation of \eqref{MM1} more directly related to the string action \eqref{Sch} in the classical limit was proposed in  \cite{Fayyazuddin:1997yf}, 
\be\label{MM2}
S_{{\tt MM2}}[A,\psi,\phi] =  \texttt{Str} \, \left(\alpha\left(-\frac{1}{4} \phi^{-1} [A^I, A^J] [A_I, A_J]  -\frac{1}{2} \,\bar{\psi}\gamma_I [A^I\,, \psi]\right)+\beta \phi \, \right), \qquad I,J=0,1,..., {D}-1,
\ee
where now $\phi(\sigma)$ is the matrix associated to $\sqrt{-g}$.
The equations of motion obtained from \eqref{MM2} are,
\be \label{MM2eq}
\begin{array}{c}
[Y^J, \{ \phi^{-1},[Y_I, Y_J] \}]=0\,, \qquad \gamma_I [Y^I\,, \psi]=0,\\[5pt]
-\frac{\alpha}{4} \phi^{-1}  [Y^I,Y^J] [Y_I, Y_J] \phi^{-1}+\beta=0,
\end{array}
\ee
where $\{ X ,Y\} :=X  Y + Y X$, is the anti-commutator. The equations of motion of the IKKT model \eqref{MM1} are obtained when $\phi=$ constant. 

The matrix models \eqref{MM1} and \eqref{MM2} are in particular related to string theory but more generally, matrix models can be regarded as describing subspaces of non-abelian algebras (of functions in non-commutative spaces) by means of constraints obtained from an action principle. The action of a generic matrix model has the form, 
\be \label{SY}
S[Y]= \langle L[Y(y)] \rangle , 
\ee
where $\langle \, \cdot \, \rangle$ is the (super)trace operation,  $L$ is a functional of functions $Y(y)$ defined on the basis of the algebra ${\cal Y} \ni y$.  From the variation of $Y$, $\delta S[Y(y)]= \langle \delta Y  C[Y(y)] \rangle = 0$ produces the equation of motion,  
\be \label{CY0}
C[Y(y)]=0. 
\ee
It is easy to show that  this equation enjoy the transformation symmetry 
\be \nonumber
\tilde{Y}= g(y)Yg^{-1}(y),
\ee 
where $g(y)$ and its inverse $g^{-1}(y)$ are functions of the generators $y$'s.  It is enough for this purpose to asume a polynomial form of the constraints \eqref{CY0}.

\subsection{From matrix models to higher spin gravity}

Matrix models can be extended to fiber bundles locally equivalent to $\cal{M} \times {\cal U(Y)} $,  with local sections given by a set of functions $Y(y;x)$, i.e. which at points $x=\{x^0, x^1,...,x^{ D-1} \}\in {\cal M}$ are expanded in the basis of an associative algebra ${\cal U(Y)} $ constructed from { polynomial functions of the generators of a Lie (super)algebra  ${\cal Y} $.}  If the representation of the algebra ${\cal Y} $ is given in terms of finite dimensional matrices then ${\cal U(Y)} $ will consists of a general linear algebra or some of its associative subalgebras. For more general representations, including infinite dimensional ones,  ${\cal U(Y)} $ will be equivalent to the universal enveloping algebra of   ${\cal Y} $. We shall focus in the latter case. This is, the local sections can be expanded as,
\be \label{Yexpand}
Y(y;x)=\sum_0^\infty \frac{1}{n!}Y^{\a(n)}(x)y_{\a(n)}, 
\ee
where $y_\a\in {\cal Y}$ is an element of the Lie (super-)algebra ${\cal Y}$ and their symmetric products $y_{\a(n)} := y_{(\a_1} \cdots y_{\a_n)} \in {\cal U(Y)}$ are given in Weyl order, for definitness, and the sum is up to infinity. Note that considering formal Taylor expansions of the sections $Y(y;x)$ in terms of the basis generators $y$'s we can extend the universal enveloping algebra $ {\cal U(Y)}$ to non-polynomial classes of functions/distributions { (see \cite{Boulanger:2013naa,Boulanger:2015uha} for their use in fractional spin gravity).} If $y$'s are finite dimensional the expansion \eqref{Yexpand} will be truncated. 

The reader familiarized with HSGR will note already a similarity of \eqref{Yexpand} with the fields of HSGR. Actually, their equivalence is up to determination of the algebra ${\cal U(Y)}$, which for higher spin gravity { it is typically  a Weyl algebra of multiple oscillators and their products with Clifford algebras \cite{Vasiliev:1990en,Vasiliev:1992av,Vasiliev:1999ba,Vasiliev:2004cp,Vasiliev:2014vwa}.} As in HSGR, we shall require local invariance of the matrix model in  $\cal{M}$-space, i.e. that the systems of constraints  \eqref{CY0} must be constant with respect to a covariant derivative. To satisfy this requirement we have to introduce a gauge connection, say $W$, and a constant curvature equation. The complete (integrable) system of constraints is given therefore by
\bea 
 \hbox{(kinematic constraints)}\qquad  &F_W=0, \qquad D_WY=0,& \label{kin}\\
\hbox{(rigid/MM constraints)}\qquad  &C(Y)=0,& \label{rig}
\eea
where $F_W:=dW+W \wedge W$ is the curvature of the gauge connection (one-form) $W$ and $Y$ should be regarded now as a zero-form in the cotangent space of ${\cal M}$, and $D_W=d+W$ is the covariant derivative, with exterior derivative $d=dx^\mu \partial_\mu$.
These equations can be split in two subsets,  the ``kinematic" ones \eqref{kin} which involves spacetime  (exterior) derivatives $d$,  and the ``rigid'' ones \eqref{rig}  consisting of the algebraic constraints coming from the matrix model \eqref{SY}.  The system \eqref{kin}-\eqref{rig} is integrable, i.e. it is closed under repeated action of the covariant derivative since, $D_WC(Y)=0$ as consequence of $D_WY=0$ and $D_WD_WY=[F_W,Y]=0$ as consequence of $F_W=0$, etc.  
The system \eqref{kin}-\eqref{rig} is also gauge invariant under infinitesimal transformations,
\be \nonumber
\delta W= D\epsilon,\qquad \delta Y= [Y,\epsilon].
\ee
Solving the kinematic equations can be done absorbing the spacetime dependency in gauge-group elements $g$ such that:
\be \label{gf} 
W=gdg^{-1},\qquad Y=gY_og^{-1},
\ee
where $Y_o$ is a spacetime-independent gauge-algebra element. The rigid constraint \eqref{rig} is left invariant so $Y_o$ must satisfy 
\be \label{CYo0} 
C(Y_o)=0.
\ee 
Thus the rigid constraints encode most of the dynamics of the system \eqref{kin}-\eqref{rig}.

As we mentioned, the equations \eqref{kin}-\eqref{rig} look quite similar to Vasiliev's HSGR equations (see an alternative form of Vasiliev's equations in \cite{Alkalaev:2014nsa}). The differences are in the details, the algebras involved and the explicit form of the constraints \eqref{rig}.  Another important detail is the preference for the use of phase-space (deformation) quantization techniques \cite{Bayen:1977ha,Bayen:1977hb}, i.e. the use of classical functions endowed with a $\*$-product to construct the associative (non-commutative) enveloping algebras ${\cal U(Y)}$, instead of working with matrices or Hilbert-space operators (see the review  \cite{Vasiliev:1999ba} for further details). Though in this section we have not referred explicitly to the $\*$-product the reader may assume that the non-commutative algebras ${\cal Y}$ and ${\cal U(Y)}$ are constructed in terms of classical functions and a $\*$-product. 

The fusion of HSGR and MMs is given by \eqref{kin}-\eqref{rig}. This means, while the constraints \eqref{rig} encode the dynamics of a MM the kinetic extensions \eqref{kin} will describe HSGR dynamics, i.e. the emergence of interacting generalized Lorentz connections ($W$) and matter fields ($Y$)  with arbitrary spin which will extend standard gauge gravity. 

Supporting these ideas, in reference \cite{Prokushkin:1999gc} it was noticed that the physical degrees of freedom of higher spin gravity coupled to matter fields in $2+1$ dimensions are encoded in a coordinates-free action of the rigid type \eqref{rig}, which is actually a matrix model equivalent to a Yang-Mills theory in even non-commutative dimensions. Actually, the equations of higher spin gravity in three dimensions are given by \cite{Prokushkin:1998vn,Prokushkin:1998bq}:
\bea
 &F_W=0, \quad D_WB=0, \quad D_WS_\a=0,& \label{Vkin}\\
&S_\a S^\a+2i(1+B)=0, \quad S_\a B+ B S_\a=0,& \label{Vrig}
\eea
where compared to \eqref{kin}-\eqref{rig} the kinetic constraints are given by  \eqref{Vkin} and the rigid ones by \eqref{Vrig}, while the zero forms are given by $Y=\{ B, S_\a\}$. Here $\a$'s are spinor indices in three dimensions. Now, according to the authors  \cite{Prokushkin:1999gc} the system  \eqref{Vkin}-\eqref{Vrig} can be reduced, using the gauge-function method \eqref{gf}, to the spacetime-coordinates-free equations \eqref{CYo0} now given by the deformed oscillator algebra \cite{Wigner:50,Yang:51},
\be \label{DHA0}
s_\a s^\a+2i(1+b)=0, \quad s_\a b+ b s_\a=0,
\ee
related to \eqref{Vrig} up to spacetime dependent similarity trasformation \eqref{gf}, $S_\a=gs_ag^{-1}$, $B=gbg^{-1}$.
Another form of writing \eqref{DHA0} is,
\be \nonumber
{ [q,p]=i(1+\nu k), \quad \{q,k \}=\{p,k \}=0,}
\ee
{which for  $k^2=1$  becomes equivalent to the Wigner deformed oscillator algebra \cite{Wigner:50}.} Here $q$ and $p$ are the coordinate and the conjugated momentum of a deformed oscillator and the $b$ field has been factorized in the product of a scalar $\nu$ and the Klein operator $k$.  { It is worth mentioning that  \eqref{DHA0} implies $osp(1|2)$ (anti-)commutation relations with $sp(2)=\{q^2,p^2,qp+pq\}$ and supercharges $q$ and $p$.  This can be extended to $osp(2|2)$ treating $k$ as an internal $u(1)$ generator and $-ikq$ and $-ikp$ as additional supercharges.} 
The matrix model action proposed in  \cite{Prokushkin:1999gc} is given by,
\be\label{PSVMM}
S_{\tt PSV}[s,b]=\langle i s_\a s^\a b - 2b -b^2 \rangle,
\ee
from which \eqref{DHA0} can be obtained variating $b$ (even degree) and $s_\a$ (odd degree) and using the correspondent (anti-)cyclic property of the supertrace $\langle \cdot \rangle$. Thus, the equations of higher spin gravity in three dimensions are somehow a covariantization in $\cal M$ space of the { Prokushkin-Segal-Vasiliev (PSV)  matrix model \eqref{PSVMM}}. 
 { The emergence of higher spin fields on non-commutative geometries has been also observed in references \cite{Sperling:2017dts,Sperling:2017gmy}.}

More generally we shall refer as to  \textit{higher-spin--matrix models} to the extension of matrix models and HSGR which can be described by the system of constraints
\bea
\begin{array}{c} 
F_W=0, \qquad D_WY=0,\\
\delta S_{\tt MM} [Y]=0,
\end{array} \label{HSMM}
\eea
where $\delta S_{\tt MM} [Y]=0$, the rigid component of the equations of motion, are obtained from the variation of  the action of  the matrix model $S_{\tt MM} [Y]$. { We say that the higher spin matrix model \eqref{HSMM} has the same type of $S_{\tt MM} [Y]$.}

\section{Deformation quantization of type IIB strings}

In  the former section we omitted explicit reference to the $\*$-product. To transform our formulas into a $\*$-product form we shall assume that for any two functions $F(y;x)$ and $G(y;x)$ of commutative variables $x$ and non-commutative variables $y$, which can be decomposed as 
$$
F:=f(x)g(y)=g(y)f(x), \qquad F':=f'(x)g'(y)=g'(y)f'(x),
$$
where $f(x)$ and $f'(x)$ are differential forms in $\cal M$-space and $g(y)$ and $g'(y)$ are elements of the algebra ${\cal U(Y)}$,  the product $F(y;x)G(y;x)$ has inserted a $\*$-product such that
$$
F(y;x)G(y;x):= (f(x)f'(x)) (g(y)\* g'(y))= (g(y)\* g'(y)) (f(x) f'(x)).
$$
Let us define now the Groenewold-Moyal  $\*$-product,
\be 
 f(y)\* h({y})~=~\int \frac{d^{2}{\xi} \, d^{2}{\zeta}}{(\pi{\theta})^{2}} \, \exp\Big( -\tfrac{2i}{{\theta}}\, (\xi^1 \zeta^2-\xi^2 \zeta^1 )\Big) \, f({y}+\xi )\, h({y}+\zeta)\,, \nonumber
\ee
for two arbitrary functions on the world-sheet $f(y)$ and $g(y)$.  The auxiliary variables $\xi$ and $\zeta$ are also world-sheet type and the quantization (Planck) constant ${\theta}$ (with units of area),  is the parameter of deformation from the classical juxtaposition product  $f(y) h(y)= h(y)f(y)$.
For example the $\*$-product of world-sheet coordinates is,
\be 
y^\a \* y^\b=y^\a y^\b +i\frac{\theta}{2}\epsilon ^ {\a\b}, \nonumber
\ee
where $y^\a y^\b=y^\b y^\a$ is the classical part of the product and the quantum deformation is given by the product of $\theta$ and the epsilon tensor $\epsilon ^ {12}=- \epsilon ^ {21}=1$. 
The $\*$-commutator is given by
\be 
[ f(y), h({y})]_{\*}:= f(y)\* h(y) -   f(y)\* h(y),\nonumber
\ee
and it is simple to show that in the classical limit
\be 
\lim_{\theta \rightarrow 0} \frac{[X, Y]_\*}{i\theta}=[X, Y]_P,\nonumber
\ee
and that
\be 
\lim_{\theta \rightarrow 0} X\* Y=XY=Y X.\nonumber
\ee

Now let us deform the string actions { \eqref{Sch} and \eqref{SNG}}. This is simple since these action principles can be expressed in terms of the Poisson brackets. The deformation-quantization version of the Nambu-Goto action is obtained substituting  $[X, Y]_P$ by $[X, Y]_\*/{i\theta}$, so that
\be \nonumber 
S_{\tt NG}^\theta[Y,\psi] = -T  \int  d^2y  \,\left(\sqrt[\*]{\frac{1}{2\theta^2 } [Y^I,Y^J]_{\*} \, \* [Y_I, Y_J]_{\*}} +\frac{2}{\theta } \bar{\psi}\* \gamma^I [Y_I\,, \psi]_{\*}\right),\nonumber
\ee
where $\sqrt[\*]{}$ is defined such that $\sqrt[\*]{f}\* \sqrt[\*]{f}= f$. Thus in the classical limit,
\be 
\lim_{\theta \rightarrow 0} S_{\tt NG}^\theta[Y,\psi]=S_{\tt NG}[Y,\psi]. \nonumber
\ee

Deforming the Schild action \eqref{Sch} results in
\be  \label{Sch*}
S^\theta_{\tt Sch}[Y,\psi ,\phi ] = -T \int  d^2y  \,\left( \frac{1}{\theta^2} \phi^{-1} \*\, [Y^I,Y^J]_\*  \*  [Y_I, Y_J]_\* +\frac{2}{\theta} \bar{\psi}\* \gamma^I [Y_I\,, \psi]_\* + \frac{1}{8} \phi \right), 
\ee
where we have used the values \eqref{alphabeta} and $\phi^{-1}$ is such that $\phi \* \phi^{-1}=1=\phi^{-1} \* \phi$. In an operator representation, which can be achieved by means of a Wigner map, the equivalent of the action \eqref{Sch*} is given by \eqref{MM2}. 
In the classical limit,
\be 
\lim_{\theta \rightarrow 0} S_{\tt Sch}^\theta[Y,\psi,\phi]=S_{\tt Sch}[Y,\psi,\phi].\nonumber %
\ee

The authors of \cite{Ishibashi:1996xs,Fayyazuddin:1997yf} assume that the Poisson bracket produced in the  classical limit of the respective matrix models is defined in a two dimensional phase space (the worldsheet). This condition may be relaxed since there is no data of the phase space dimension in the matrix model. { In the deformation quantization approach we can declare  the dimension of the classical phase space in the action itself,}
\be\label{Sch*2d}
S_{\texttt{IIB}}^\theta [Y,\psi] = -{\cal T} \int  d^{2d} y  \,\left( \frac{1}{\theta^2} \phi^{-1} \*\, [Y^I,Y^J]_\*  \*  [Y_I, Y_J]_\* +\frac{2}{\theta} \bar{\psi}\* \gamma^I [Y_I\,, \psi]_\* + \frac{1}{8} \phi \right), 
\ee
{ as an example of generalization of the model \cite{Ishibashi:1996xs,Fayyazuddin:1997yf}.} Now the tension ${\cal T}$  has been rescaled according to the choice of the dimensions of the phase space. Now in the classical limit what is produced is a theory of { extended} objects. In this case the $\*$-product will be given by
\be \label{f*g2d}
 f(y)\* g(y)~=~\int \frac{d^{2d}{\xi} \, d^{2d}{\zeta}}{(\pi{\theta})^{2d}} \, \exp\Big( -\tfrac{2i}{{\theta}}\, \bar{\xi} \zeta \Big) \, f(y+\xi )\, g(y+\zeta)\,, 
\ee
where $\bar{\xi} \zeta = \xi^{\a} \zeta_{\a}$
and we have used the symplectic matrix 
\begin{eqnarray}
C_{\a \b}=C^{\a \b}:=
\left(%
\begin{array}{cc}
  0 & \mathds{1}_{d\times d} \\
  -\mathds{1}_{d\times d} & 0 \\
\end{array}
 \right),\nonumber
\end{eqnarray}
to rise or lower the phase space index $\bar{\xi}=\xi^{\a}$ according to the following conventions,
\be \nonumber
\xi^\a = C^{\a \b} \xi_\b \, , \qquad \xi_\a = \xi^\b C_{\b \a}\, , \quad \hbox{where}\quad C_{\a \b} = C^{\a \b}\,.
\ee 
Thus the $\*$-product of two vectors in phase space yields, 
\be \nonumber \label{y*y}
y^\a \* y^\b =y^\a  y^\b + \frac{i {\theta}}{2} C^{\a \b}\,.
\ee 

The equations of motion obtained from \eqref{Sch*2d} are,
\be \label{eomMM2}
\begin{array}{c}
[Y^J, \{ \phi^{-1},[Y_I, Y_J]_\* \}_\*]_\*=0\,, \qquad \gamma_I [Y^I\,, \psi]_\*=0,\\[5pt]
\phi^{-1} \* [Y^I,Y^J]_\*\* [Y_I, Y_J]_\* \* \phi^{-1}-\frac{1}{8}=0,
\end{array}
\ee
where $\{ X ,Y\}_\*:=X\* Y + Y\*X$, is the $\*$--anti-commutator. The analogous to the IKKT model \eqref{MM1} is obtained when $\phi=1$ and $d=1$  in \eqref{f*g2d}.
The equations of motion are obtained from the property
$$
\int d^{2d}y  A\* B = (-1)^{|A||B|} \int d^{2d}y  B\* A,
$$
for Grassmann parity $|A|=0,1$.

\section{Type IIB higher spin matrix models in $D=2,3,4\, ~ \texttt{Mod} ~ 8 $ }

Let us construct a specific model as an illustration of the ideas presented here. Now making explicit reference to the star-product \eqref{f*g2d} the equations of motion are:
\bea
&d\bW+\bW\* \wedge \bW=0,\label{F0}&\\
&d\bY^I+[\bW,\bY^I]_\*=0,\qquad d\bPsi+[\bW,\bPsi]_\*=0,\qquad d\bPhi+[\bW,\bPhi]_\*=0,&\label{D0}\\[5pt]
&[\bY^J, \{ \bPhi^{-1},[\bY_I, \bY_J]_\* \}_\*]_\*=0\,, \qquad \gamma_I [\bY^I\,, \bPsi]_\*=0&\label{r1}\\
&\bPhi^{-1} \* [\bY^I,\bY^J]_\*\* [\bY_I, \bY_J]_\* \* \bPhi^{-1}-\frac{1}{8}=0.&\label{r2}\\[5pt]
& { I \quad \rightarrow} \quad  \{ \mu[n]= [\mu_1\mu_2\cdots \mu_n]\,,\: \mu_1< \mu_2 < \cdots < \mu_n,\quad \mu_k=0,1,...,D-1\}.\label{indices}
&
\eea
The kinetic \eqref{kin} and rigid  \eqref{rig} equations are now  \eqref{F0}-\eqref{D0} and \eqref{r1}-\eqref{r2} respectively.  The rank of indices \eqref{indices} will be explained below. In this specific case, the rigid equations are obtained from the variation of the action for type IIB matrix models \eqref{Sch*2d}. To the theory \eqref{F0}-\eqref{r2} we shall refer as to type IIB higher spin matrix model.

We have not declared what is the dimension of the spacetime $D$, neither the rank of the capital indices $I$, $J$ or the algebras involved. The system above is formally integrable but the choices of the algebras involved and of the dimension of the space-time is an engineering problem, it depends of what we want to describe. In what follows we shall solve the system  \eqref{F0}-\eqref{r2} employing similar techniques than in reference \cite{Valenzuela:2015gia}.

Like in Vasiliev's HSGR, let us consider the Heisenberg algebra, 
\be \nonumber 
{\cal Y}=\{ [y_\a,y_\b]_\*=i \theta C_{\a\b} \,; \quad \a,\b=1,...,2d\},
\ee
whose universal enveloping algebra, i.e. the algebra of polynomials in  ${\cal Y}$, is given by
\be \label{UY}
{\cal U(Y)}=\{ y_{\a(n)}\,; n=0,...,\infty\},
\ee
where { the symmetric products}
\be \label{yn}
y_{\a(n)} := y_{(\a_1}\*  y_{\a_2}\* \cdots \* y_{\a_n)}= y_{\a_1}  y_{\a_2}\cdots y_{\a_n},
\ee
{are by the properties of the $\*$-product equivalent to the classical monomial in the right hand side.}
In \eqref{yn}  we have used standard notations for the symmetrization of the products, with the factorial normalization. 
Now the fields \eqref{F0}-\eqref{r2} can be expanded in the basis  of ${\cal U(Y)}$ as follows
\bea 
&\bW(y;x)=\sum\limits_{n,\a,} \frac{1}{n! }W^{\a(n)}(x)y_{\a(n)}, \label{Wn}&\\
&\bY^I(y;x)=\sum\limits_{n,\a} \frac{1}{n!}Y^{I,\a(n)}(x)y_{\a(n)} ,\label{Yn}& \\
&{ \bPsi(y;x)=\sum\limits_{n,\a}\frac{1}{n!}\psi^{\a(n)}(x)y_{\a(n)}, }\label{Psin}&\\
&{ \bPhi(y;x)=\sum\limits_{n,\a}\frac{1}{n!}\phi^{\a(n)}(x)y_{\a(n)} },& \label{Phin}\\
&n=0,...,\infty,\quad \a=1,...,2d. &\nonumber
\eea
With respect to the $\*$-commutator product $[\cdot ,\cdot ]_\*$  the second order polynomials $y_{\a(2)} $ generate a representation of the $sp(2d)$ algebra, { which for $2d=2^{[D/2]}$} contains a representation of the anti de Sitter algebra $so(D-1,2)$. The latter algebra can be used to make explicit the Lorentz covariance of the system \eqref{F0}-\eqref{r2}. Indeed the variables $y_{\a(n)}$ transform in the spin $n/2$ adjoint representation of the Lorentz algebra in $D$ spacetime dimensions. From this observation it is clear that $\bW$ contains at level $n=2$ the $AdS_D$ Lorentz connection { and for $n\neq 2$ these are higher spin gravity gauge fields, which justify the ``higher spin gravity'' part of the title of this paper.} This is, the $AdS_D$ space is a natural solution of the system of equations, if we put all the remaining fields to zero, for example. Now the spacetime dimension is specified to be $D$, according to the choice of the algebra \eqref{UY}.

{ Now we should clarify the meaning of the indices \eqref{indices}. The rigid equations \eqref{r1}-\eqref{r2} are written in a form that reminds us the type IIB matrix model field equations  \eqref{MM2eq}-\eqref{eomMM2}, but as we shall justify, the fields involved admit more general labels, i.e. with $I$ labeling target space Lorentz multivectors instead of just vectors. As it was shown in  \cite{Valenzuela:2015gia}, the phase space monomials $y_{\a(2)} $ ($\a=1,...,2^{[D/2]}$), in the classical level, parametrize an algebraic  variety (say $\bf M$) whose coordinates are conveniently labeled by Lorentz multivectors in $D$ dimensions. $\bf M$ admits a covariant non-commutative deformation (quantization), i.e. introducing the $\*$-product \eqref{f*g2d} in the algebra of functions in the space $\bf M$, which is Lorentz covariant. Surprisingly enough, the coordinates of the non-commutative version of $\bf M$ will satisfy the constraints \eqref{r1}-\eqref{r2}. {\it A posteriori} this result justifies the use of multivector labels. Thus we can change the notation $\bY^I$ by  $\bY^{[\mu_1 \mu_2 \cdots \mu_n]}=:\bY^{\mu[n]}$,   $\mu_k=0,...,D-1$, for some values of $n$ to be specified below. } In order to observe this let us introduce the exterior algebra of real (Majorana) Dirac matrices,
\be \label{Gamman}
\gamma^{\bmu[n]}:=\gamma^{[\bmu_1} \cdots \gamma^{\bmu_n]}, \qquad \mu=0,...,D-1,
\ee
constructed from the Clifford algebra $\{ \gamma_\mu \,, \gamma_\nu \}= 2\ {\eta}_{\mu \nu}$, with $\texttt{diag}({\eta}_{\mu \nu})= (-1,1,\cdots ,1).$ 
 Here $n$ is in the set of integers $1,2 ~ \texttt{Mod} ~ 4 $ and $n \leq D$ for $D$ even or  $n\leq [D/2]$ for $D$ odd (see \cite{vanHolten:1982mx}), for which the $\gamma^{\bmu[n]}$ matrices are independent.
The matrices \eqref{Gamman} span a representation of the $sp(2^{[D/2]})$ algebra, which acts on the space of $2d=2^{[D/2]}$-components spinors.  The symmetric monomials $y_{\a(2)} $ also span a representation of $sp(2^{[D/2]})$,  with respect to the $[\cdot,\cdot]_\*$ product, and indeed they can be given multivector labels using the bi-linear combinations provided by \eqref{Gamman}. This is
\be \label{X}
X^{\mu[n]}=\frac{1}{4} \bar{y}\gamma^{\mu[n]}  y ,
\ee
are $sp(2^{[D/2]})$ generators.
 
{ To make more explicit the correspondence \eqref{indices} $I \leftrightarrow \mu[n]$ let} us split the interval $ I\, \in \, \{1,...,\texttt{dim} (sp(2^{[D/2]}))\}$ in subspaces 
\be 
I \in \oplus_{n \in 1,2\,\texttt{Mod}\, 4} I_{[n]}, \qquad n=1,2~ \texttt{ Mod } 4, \nonumber
\ee
where $I_{[n]}$ are intervals of integer numbers of size $\binom{D}{n}$, in correspondence with the independent components of multivectors $X^{\mu[n]}$.  To each element in the subset  $ I_{[n]}$ we assign a single multivector index
 \be \nonumber 
 I \in I_{[n]} \quad \rightarrow \quad  \mu[n]= [\mu_1\mu_2\cdots \mu_n]\,,\quad \mu_1< \mu_2 < \cdots < \mu_n,\quad
 \mu=0,1,\cdots D-1.
 \ee
 For example we have the correspondence,
\bea
&&\hspace{2mm}D=2+1:\:  I_{[1]}=\{1,2,3\} \rightarrow \mu=0,1,2,\:I_{[2]}={4,5,6} \rightarrow \mu[2]=[01],[02],[12],\label{D3}\\
&&\begin{array}{l}
D=3+1:\:  I_{[1]} = \{1,...,4\} \rightarrow \mu=0,1,2,3,\: \\
\hspace{19mm}I_{[2]}=\{5,...,10\} \rightarrow \mu[2]=[01],[02],[03],[12],[13],[23],
\end{array} \nonumber 
\eea
in the respective space-time dimensions.
 We can write now
 \be \nonumber%
 \bY^{\mu[n]}(y;x)=\sum_{m,\a} \frac{1}{m!}Y^{{\mu[n]},\a(m)}(x)y_{\a(m)} ,
 \ee
 instead of \eqref{Yn}, and 
 \be  \nonumber%
 d\bY^{\mu[n]}+[\bW,\bY^{\mu[n]}]_\*=0,
 \ee
 instead of $d\bY^I+[\bW,\bY^I]_\*=0$ in \eqref{D0}. Now the meaning of the rank of indices \eqref{indices} is understood but we need to specify how do we construct the invariants $X^IX_I = X^IX^J  \eta_{IJ}$. Indeed, $X^IX_I$ is a $sp(2^{[D/2]})$ invariant. In multivector notations the $sp$-Killing-metric looks like
\be 
\texttt{sign}\, \eta_{IJ}=\texttt{sign}\, \eta_{\mu[n]_< \nu[n]_<} =(-1)^{[n/2]}\texttt{sign}\,\eta_{\mu_1 \nu_1}\eta_{\mu_2 \nu_2}\dots \eta_{\mu_n \nu_n}\, \,, \label{etaIJ}
\ee
where the tensor
\be 
\eta_{{\bmu[n]}\, {\bnu[n]} }=\frac{(-1)^{[n/2]}}{(n!)^2} \eta_{\bmu_1 \brho_1} \eta_{\bmu_2 \brho_2} \cdots \eta_{\bmu_n \brho_n} \delta_{\bnu_1 \bnu_2 \cdots \bnu_n}^{\brho_1 \brho_2 \cdots \brho_n}, \nonumber
\ee
is constructed from the product of the Lorentz metric tensor $\eta_{\bmu \brho}$, and the indices $\mu[n]_<=[\mu_1\mu_2... \mu_n]$ are ordered,  $\mu_1<\mu_2<... <\mu_n$.

Now, the rigid constraints \eqref{r1}-\eqref{r2} look like
{
\bea
&\sum\limits_{m=1,2\,\, \texttt{Mod} 4}'  \frac{(-1)^{[m/2]}}{m!} [\bY^{\mu[m]}, \{ \Phi^{-1},[\bY_{\nu[n]}, \bY_{\mu[m]}]_\* \}_\*]_\*=0\,,&\label{rr1}\\ 
&\sum\limits_{ m=1,2\,\, \texttt{Mod} 4 }'  \frac{(-1)^{[m/2]}}{m!}  \gamma_{\mu[m]} [\bY^{\mu[m]}\,, \bPsi]_\*=0&\label{rr2}\\
& \, \sum\limits_{m,n=1,2\,\, \texttt{Mod} 4}'  \frac{(-1)^{[m/2]+[n/2]}}{m!n!} \bPhi^{-1} \* [\bY^{\mu[m]},\bY^{\nu[n]}]_\*\* [\bY_{\mu[m]}, \bY_{\nu[n]}]_\* \* \bPhi^{-1}-\frac{1}{8}=0,&\label{rr3}
\eea
} where $\sum'$ means to count in the interval of integers $n$ for which the $\gamma^{\bmu[n]}$'s span the basis of independent matrices \eqref{Gamman} (see comment below \eqref{Gamman}).

\subsection{Some solutions}\label{solution:sec}
 
Let us show that 
\be \label{sol}
{ \bY^{\bmu[n]}=X^{\bmu[n]},\quad  \bPsi=0,\qquad  \bPhi=\texttt{constant},}
\ee 
where $X^{\bmu[n]}$ is given by  \eqref{X}, solve the systems of equations \eqref{rr1}-\eqref{rr3}. It is trivial that  $\Psi=0$ solves  \eqref{rr2}. If $\Phi$ is constant then  \eqref{rr1} is reduced to,
\be \nonumber
\sum\limits_{m=1,2\,\, \texttt{Mod} 4}'  \frac{(-1)^{[m/2]}}{m!} [\bY^{\mu[m]}, [\bY_{\nu[n]}, \bY_{\mu[m]}]_\*]_\*=0. 
\ee 
That this equation is solved by \eqref{X} was shown in reference \cite{Valenzuela:2015gia}. For that purpose we can use the identity, 
 \be\nonumber
C_{\a \b} C_{\a' \b'} +C_{\a' \b} C_{\a \b'} =\frac{1}{2^{[D/2]-1}} \sum _{m=1,2\,\, ~ \texttt{Mod} ~ 4}' \frac{(-1)^{[m/2]+1}}{m!} (\gamma^{\nu[m]})_{\a \a'}  (\gamma_{\nu[m]})_{\b \b'},
\ee
and that $y^\a y_\a=0$. 

Similarly to \eqref{rr1}, using \eqref{sol} and
\begin{eqnarray} \nonumber
y_\bfa y_\bfb \* y_\bfxi y_\bfz = y_\bfa y_\bfb  y_\bfxi y_\bfz +i \frac{{\theta}}{2}\left(
C_{\bfa \bfxi} y_\bfb y_\bfz + C_{\bfa \bfz} y_\bfb y_\bfxi + C_{\bfb \bfxi} y_\bfa y_\bfz + C_{\bfb \bfz} y_\bfa y_\bfxi 
\right)  - \frac{{\theta}^2}{4}\left(
C_{\bfa \bfxi} C_{\bfb \bfz}+C_{\bfa \bfz} C_{\bfb \bfxi}   \right),\nonumber
\end{eqnarray}
 we can compute the value of
\be 
 \sum\limits_{m,n=1,2\,\, \texttt{Mod} 4}'  \frac{(-1)^{[m/2]+[n/2]}}{m!n!}  [\bY^{\mu[m]},\bY^{\nu[n]}]_\*\* [\bY_{\mu[m]}, \bY_{\nu[n]}]_\* \* = 2\theta^4 2^{2 [D/2]} (1+2^{2 [D/2]}).\nonumber
 \ee
Hence, from \eqref{rr3} we obtain
\be
{ \bPhi = \pm \theta^2 2^{[D/2]+2} \sqrt{1+2^{[D/2]}}.}\nonumber
\ee 
We can extend these solutions to solutions of the whole system \eqref{F0}-\eqref{indices} using gauge functions 
\be
g= \exp \left( \epsilon_{{\nu[n]}} (x)  X^{\nu[n]} \right), \nonumber%
\ee
where $\epsilon_{{\nu[n]}} (x)$ are functions of the point $x$ of the spacetime, so that
\be \nonumber
{\bW= g\*dg^{-1}, \qquad \bY^{\bmu[n]}=g\*X^{\bmu[n]}\*g^{-1},\quad  \bPsi=0,\qquad  \bPhi=  \pm \theta^2 2^{[D/2]+2} \sqrt{1+2^{[D/2]}}.}
\ee 

As we observed, we can unify multivector type of matrix models and higher spin gravity.

\section{Conclusions}

We have presented a general class higher spin gravity models which incorporate matrix models in their own definition. The techniques are similar to standard Vasiliev's higher spin gravity, the difference is in the internal algebras and in part of the equations of motion which now are derived from matrix models. These theories put in closer contact higher spin gravity and string theory, as according to e.g. \cite{Banks:1996vh,Ishibashi:1996xs,Fayyazuddin:1997yf}, matrix models are candidates for non-perturbative theories of strings and M-theory. We constructed and provided solutions for a particular model related to  type IIB strings, and therefore called { {\it type IIB higher spin gravity}}. The matrix models constructed here extend those found in \cite{Valenzuela:2015gia} in that now spacetime dimensions are added. An interesting aspect of these matrix models is that they incorporate coordinates of extended objects, as multivector coordinates of rank $k$ are related to $k$-dimensional objects.  For instance,  part of the  rigid equations of motion for (multivector) matrix model in $3+1$ dimensions (without fermions and for a constant scalar field) can be reduced to the form,
\bea 
[\bY^\bnu,[\bY_\bmu,\bY_\bnu]_\*]_\*- \frac{1}{2} 
[\bY^{\bnu\blambda},[\bY_\bmu,\bY_{\bnu\blambda}]_\*]_\*&=& 0\,, \nonumber
\\[5pt]
[\bY^\bnu,[\bY_{\bmu\brho},\bY_\bnu]_\*]_\*- \frac{1}{2} 
[\bY^{\bnu\blambda},[\bY_{\bmu\brho},\bY_{\bnu\blambda}]_\*]_\* &=& 0\,. \nonumber
\eea
Some solutions of this system contain Plucker coordinates of planes through $3+1$ dimensions as shown in \cite{Valenzuela:2015gia}. An interesting problem to address is whether the classical limit of the respective matrix model  \eqref{F0}-\eqref{r2}, or \eqref{F0}, \eqref{D0}, \eqref{rr1}-\eqref{rr3}, reproduces Polyakov's string theory in $3+1$ dimensions with fine structure \cite{Polyakov:1986cs}. 

{ Note that in this paper we have not studied the details of the physical degrees of freedom contained in the constructed models. This can be done using similar methods than in matrix models and higher spin gravity, i.e. by means of perturbation theory around some solutions of the models, for instance the ones provided in section \ref{solution:sec}.}

{ We would like to mention that} multivector extensions of spacetime also  appear in the context of E11-type string theories and supergravities \cite{Riccioni:2007ni,West:2012qz,West:2016xro} and in the study of higher spin fields dynamics in the approach \cite{Plyushchay:2003tj,Bandos:2004nn,Bandos:2005mb}. These theories and ours suggest the existence of mutivector extended spacetimes which may be a necessary, or just convenient, for the formulation of quantum theories of gravity.

We expect that our proposal will be useful to  find deeper connections between string theories, M-theory and higher spin gravity. We encourage interested readers to construct and study in detail  specific models falling in the category here presented.

\subsection*{Acknowledgments}

I would like to thank to the organizers of the {\it Workshop on Higher Spin Gauge Theories}, held in UMONS on April 2017, Thomas Basile, Roberto Bonezzi, Nicolas Boulanger, Andrea Campoleoni, David De Filippi and 
Lucas Traina, for their kind invitation and the opportunity to present the ideas here exposed. {  I would like to thank also  Nicolas Boulanger, Patricia Ritter, Evgeny Skvortsov and Per Sundell for their insightful comments. }



\end{document}